%% file: thermocellpaper2.tex
\RequirePackage{amsmath}
\documentclass[12pt]{iopart}
\usepackage{graphicx}
\usepackage{amsmath}
\usepackage{amsfonts}
\usepackage{mathtools}
\usepackage{amssymb}
\usepackage[colorlinks=true,linkcolor=blue,citecolor=blue,filecolor=cyan,urlcolor=blue,breaklinks=true]{hyperref}

\usepackage{cite}
\usepackage{subfigure}
\usepackage{subfigure}

\newcommand{\mysl}{\sl}

\begin{document}

\title{Efficiency of cellular information processing}

\author{Andre C. Barato, David Hartich and Udo Seifert}
\address{II. Institut f\"ur Theoretische Physik, Universit\"at Stuttgart, 70550 Stuttgart, Germany}
\eads{\mailto{barato@theo2.physik.uni-stuttgart.de}\\\mailto{hartich@theo2.physik.uni-stuttgart.de}\\
\mailto{useifert@theo2.physik.uni-stuttgart.de}}
\begin{abstract}
We show that a rate of conditional Shannon entropy reduction, characterizing the learning of an internal process about an external process, is bounded by the thermodynamic entropy production.
This approach allows for the definition  of an informational efficiency that can be used to study cellular information processing.
We analyze three models of increasing complexity inspired by the {\mysl E. coli}
sensory network, where the external process is an external ligand concentration jumping between two values. We start with a simple model for which ATP must be consumed so that a protein inside the
cell can learn about the external concentration. With a second model for a single receptor we show that the rate at which the receptor learns 
about the  external environment can be nonzero even without any dissipation inside the cell since chemical work done by the external process compensates for this learning rate. 
The third model is more complete, also containing adaptation. For this model we show {\mysl inter alia} that a bacterium in an environment  that changes at a very slow time-scale is quite inefficient, dissipating much more than it learns.
Using the concept of a coarse-grained learning rate, we show for the model with adaptation that while the activity learns about the external signal the option of changing the methylation level
increases the concentration range for which the learning rate is substantial.
\end{abstract}%

\section{Introduction}
Biological systems must process information about fluctuating environments, with a classical 
example being a cell computing an external ligand concentration by time-averaging \cite{berg77}.
Understanding the thermodynamics of cellular information processing is of central importance 
and has attracted much interest recently. Particularly, the role of energy dissipation has been studied
for {\mysl E. coli} \cite{lan12} and eukaryotic \cite{palo13} adaptation, a cell computing an external ligand concentration \cite{meht12}, biochemical sensing \cite{qian05c,qian07,tu08a,bara13b,skog13,gove13,beck13,lang14}, and
proofreading \cite{andr08,sart13,muru12,muru14} (see also \cite{hopf74,nini75,benn79} for older works).

In these studies, energy dissipation is characterized by the familiar entropy production of 
stochastic thermodynamics \cite{seif12}. However, they do not consider an entropic rate characterizing 
information processing of a noisy external environment that is related to the thermodynamic entropy production through an inequality.
In related work \cite{bara13b}, we have shown that the rate of mutual information between an internal process, corresponding to
chemical reactions inside the cell, and an external process is {\sl not} bounded by the thermodynamic entropy production.

More broadly, the relation between information and thermodynamics  is a very active topic, with theoretical studies focused on second law 
inequalities and fluctuation relations \cite{touc00,touc04,cao09,saga10,horo10,abre11a,saga12,ito13,espo12b,mand12,deff13,bara13,bara14}, as well as experimental works \cite{toya10a,beru12}. Of particular relevance to this paper,
it has recently been realized that a bipartite Markov process provides a simple realization of a Maxwell's demon \cite{hart14}, where the entropic rate characterizing information is the entropy 
reduction rate of a subsystem due to its coupling to the other subsystem composing the bipartite system. This entropy rate has been studied for the first time by Allahverdyan et al. \cite{alla09} for two 
Brownian particles coupled to different heat baths, where it was interpreted as an information flow.  
For bipartite jump processes a closely related entropic rate, also named information flow, has been studied more recently in \cite{horo14}. Moreover, in related work
a relation between dissipation and information about an external signal has been obtained \cite{stil12}.

In this paper, we show that for a bipartite process, with  an internal process learning about an unaffected external process, the rate at which the uncertainty about 
the external environment, represented by a conditional Shannon entropy, decreases due to the dynamics of the internal process is bounded by the thermodynamic entropy production.
We call this rate the learning rate. This concept allows for the definition of an informational efficiency 
for biological information processing. Particularly, we consider three different models 
inspired by the {\mysl E. coli} chemotaxis signaling network \cite{bark97,bial12,sour12}, 
where, for simplicity, the external process is assumed to correspond to an external ligand concentration jumping between two values.     

We demonstrate how an informational efficiency can be defined and analyzed in a simple four-state model, for which the
entropy production corresponds to ATP consumption inside the cell. For this system, a comparison with the efficiency of molecular motors is possible.
We then consider a system where the internal process corresponds to an equilibrium Monod-Wyman-Changeux (MWC) \cite{mono65,marz13} model for a single {\mysl E. coli}  receptor. 
In this case, even if there is no energy consumption inside the cell, the learning rate is bounded by the chemical work done by the external process.
Finally, we analyze a more complete model for {\mysl E. coli}  receptors including adaptation \cite{lan12}. We show that  the learning rate can
even exceed  the rate at which the cell consumes the free energy source if work done by the external process compensates for it. We also analyze a coarse-grained learning rate for the model with adaptation: comparing it
with the MWC model not including adaptation, we find that the concentration range for which the learning rate is non-negligible  increases with adaptation.

The paper is organized as follows. In Sec. \ref{sec2}, we define the class of Markov processes and observables we study in this paper. Moreover, we
show that the learning rate is bounded by the thermodynamic entropy production. In Sec. \ref{sec3}, the four state model is studied,
while Sec. \ref{sec4} contains the MWC like model. In Sec. \ref{sec5}, the model with adaptation is analyzed. We conclude in Sec. \ref{sec6}.  


\section{Bipartite systems with an external process}
\label{sec2}
\subsection{Transition rates and thermodynamic entropy production}

Let us first introduce discrete bipartite Markov processes \cite{bara13b,bara13a}, where a state $(x,y)= (\alpha,i)$ is labeled by two variables. Moreover, it is assumed that
$x$ is an external process unaffected by the internal process $y$. The transition rates, from $(\alpha,i)$ to $(\beta,j)$, are defined as
\begin{equation}
w_{ij}^{\alpha\beta}\equiv\left\{
\begin{array}{ll} 
 w^{\alpha\beta} & \quad \textrm{if $i=j$ and $\alpha\neq\beta$}, \\
 w^{\alpha}_{ij} & \quad  \textrm{if $i\neq j$ and $\alpha=\beta$},\\
 0 & \quad \textrm{if $i\neq j$ and $\alpha\neq\beta$}, 
\end{array}\right.\,
\label{defrates2}
\end{equation}
where a transition changing both variables is not allowed. The fact that $x$ is an external process
implies that the transition rates $w^{\alpha\beta}$ are independent of the internal variable $i$. 
Denoting the stationary probability distribution by $P_i^{\alpha}$, the thermodynamic entropy
production is given by \cite{seif12}
\begin{equation}
\sigma= \sum_{\alpha}P^\alpha \sum_{\beta\neq\alpha} w^{\alpha\beta}\ln \frac{w^{\alpha\beta}}{w^{\beta\alpha}}+\sum_{i,\alpha}P_i^\alpha\sum_{j\neq i} w^\alpha_{ij}\ln\frac{w^\alpha_{ij}}{w^\alpha_{ji}}\equiv \sigma_x+\sigma_y\ge 0,
\label{thermoent}
\end{equation} 
where $P^\alpha= \sum_iP_i^\alpha$.  The first (second) term of the right hand side of (\ref{thermoent}), related to external (internal) jumps, is denoted by $\sigma_x$ ($\sigma_y$).

\subsection{Learning rate and informational efficiency}

In the stationary state the conditional Shannon entropy of $x$ given $y$ is 
\begin{equation}
H[x|y]= -\sum_{i,\alpha} P_i^\alpha \ln P(\alpha|i),
\end{equation}
where $P(\alpha|i)= P_i^\alpha/P_i$, with $P_i= \sum_\alpha P_i^\alpha$. This Shannon entropy  quantifies the uncertainty about the 
external process given the internal variable. Therefore, the stationary rate at which the internal process reduces the uncertainty of the external process
due to its jumps can be written as 
\begin{equation}
l_y\equiv\sum_{i,\alpha}\sum_{j\neq i}\left(P_j^\alpha w^{\alpha}_{ji}-P_i^\alpha w^{\alpha}_{ij}\right)\ln P(\alpha|i)= -\sum_{i,\alpha}P_i^\alpha \sum_{j\neq i} w^{\alpha}_{ij}\ln \frac{P^{\alpha}_i}{P^{\alpha}_j}.
\label{learnrate}
\end{equation}   
In other words, $l_y$, which we will call the learning rate, is the rate 
at which $y$ through its dynamics learns about $x$. Note that the $y$ jumps do not change the stationary Shannon entropy of $x$, therefore, the learning rate is also equal to
the rate of change in the stationary mutual information due to the $y$ jumps \cite{alla09,horo14}. The full time derivative of $-H[x|y]$ in the stationary state  reads
\begin{equation}
-\frac{{\rm d}}{{\rm d}t}H[x|y]= l_y-h_x=0,
\label{consl}
\end{equation} 
where 
\begin{equation}
h_x\equiv \sum_{i,\alpha}\sum_{\beta\neq \alpha}\left(P_i^\alpha w^{\alpha\beta}-P_i^\beta w^{\beta\alpha}\right)\ln P(\alpha|i)= \sum_{i,\alpha}P_i^\alpha \sum_{\beta\neq\alpha} w^{\alpha\beta}\ln \frac{P^{\alpha}_i}{P^{\beta}_i},
\label{eq:hx_def}
\end{equation}  
arises from the $x$ jumps. Hence, equation (\ref{consl}) implies $h_x= l_y$. 

The rates $h_x$ and $l_y$ have been recently considered in \cite{hart14}, where $h_x$, for example, is  interpreted as the rate of the entropy
reduction of subsystem $x$ due to its coupling with $y$. The conservation law $h_x=l_y$ simply means that the rate of entropy reduction of $x$
is precisely the rate with which $y$ learns about $x$. These entropic rates have also been considered recently in \cite{horo14}, where they are referred to as information flow and obtained from 
the time derivative of the mutual information. Moreover, to our knowledge the first reference to introduce similar entropic rates is \cite{alla09}, where
two coupled Langevin equations were studied and the entropic rate is introduced as a time derivative of time-delayed mutual information. 

The learning rate fulfills  $l_y\ge 0$, which comes from the fact that the transfer entropy from $y$ to $x$ is an upper bound on $-l_y$ and equal to zero if $x$ is an external process unaffected by $y$ \cite{hart14,alla09}. 
Whereas the positivity of the above thermodynamic entropy production (\ref{thermoent}) represents the second law for the full system,
it is straightforward to show that the second law for the subsystem $y$ reads $\sigma_y-l_y\ge 0$ \cite{hart14,dian14}. Moreover, since
$x$ is an external process, if we integrate out the $y$ variable $x$ is still a Markovian process, therefore $\sigma_x\ge 0$, which implies
\begin{equation}
l_y\le \sigma_y\le \sigma.
\label{mainine}
\end{equation}
This inequality is a main foundation of the present paper: for a bipartite system with $x$ being an external process,
the rate at which $y$ learns about $x$ is bounded by the thermodynamic entropy production, which characterizes the dissipation necessary for a nonzero learning rate.
From inequality (\ref{mainine}), the following informational  efficiency can be defined, 
\begin{equation}
\eta\equiv l_y/\sigma_y\le 1.
\label{mainine2}
\end{equation}
If the external process is further assumed to be in equilibrium ($\sigma_x=0$), which is the case in all the examples studied in this paper, then $\sigma=\sigma_y$ and $\eta= l_y/\sigma$.

In the following we demonstrate with three examples that the framework discussed in this section allows the study of cellular information processing, with 
the cost and thermodynamic efficiency of learning about an external random environment being well characterized. Moreover, we show that while the learning rate 
$l_y$ is bounded by $\sigma$, it can be larger than the energy consumed by the cell if the external process does work.

\subsection{Coarse-grained learning rate}

For the MWC single receptor model in Sec. \ref{sec4} and the model with adaptation in Sec. \ref{sec5}, we have to consider an internal process which comprises  two variables $y=(y_1,y_2)$.
The internal transition rates are now written as $w^{\alpha}_{ij}=w^{\alpha}_{(i_1i_2)(j_1j_2)}$. 
In this case, another relevant quantity is the coarse-grained learning rate 
\begin{equation}
l_{y_1}= -\sum_{i_1,\alpha}P_{i_1}^\alpha \sum_{j_1\neq i_1} W^{\alpha}_{i_1j_1}\ln \frac{P^{\alpha}_{i_1}}{P^{\alpha}_{j_1}},
\label{learnratecg}
\end{equation}
where 
\begin{equation}
 W^{\alpha}_{i_1j_1}\equiv \sum_{i_2,j_2}\frac{P^\alpha_{i_1 i_2}}{P^\alpha_{i_1}} w^{\alpha}_{(i_1i_2)(j_1j_2)}, 
\end{equation}
with $P^\alpha_{i_1}= \sum_{i_2}P^\alpha_{i_1i_2}$. This is the rate at which  $y_1$ reduces the conditional Shannon entropy
\begin{equation}
H[x|y_1]= -\sum_{i_1,\alpha} P_{i_1}^\alpha \ln P(\alpha|i_1),
\end{equation}
due to its jumps. Hence, it is the rate at which the coarse-grained internal process $y_1$ learns about $x$.
The contribution due to $x$ jumps to $-\frac{\rm d}{{\rm d}t}H[x|y_1]=0$ is
\begin{equation}
h^{(1)}_x\equiv \sum_{i_1,\alpha}P_{i_1}^\alpha \sum_{\beta\neq\alpha} w^{\alpha\beta}\ln \frac{P^{\alpha}_{i_1}}{P^{\beta}_{i_1}},
\label{hxcg}
\end{equation}
and the conservation law for the coarse-grained learning rate reads $h^{(1)}_x= l_{y_1}$.
From the log sum inequality, we obtain $l_{y_1}\le l_y$, showing that the coarse-grained variable $y_1$ cannot learn more than the full variable $y$.
Moreover, the coarse-grained entropy production is defined as \cite{espo12} 
\begin{equation}
\tilde{\sigma}_{y_1}= \sum_{\alpha,i_1}\sum_{j_1\neq i_1} P_{i_1}^\alpha W^\alpha_{i_1j_1}\ln\frac{W^\alpha_{i_1j_1}}{W^\alpha_{j_1i_1}},
\label{thermoentcg}
\end{equation}
which provides a lower bound on $\sigma_y$. Finally, similar to the second law inequality for a subsystem we also have  
\begin{equation}
l_{y_1}\le \tilde{\sigma}_{y_1}.
\end{equation}
We point out that this coarse-grained learning rate and the above discussion about it are novel.

\section{Toy model}
\label{sec3}

\begin{figure}%
 \centering%
 \subfigure[]{\includegraphics{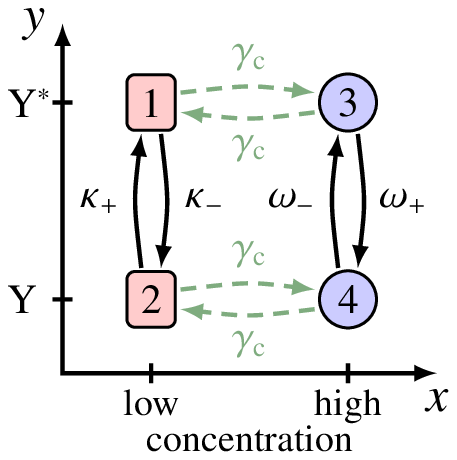}\label{fig1a}}%
 \subfigure[]{\includegraphics{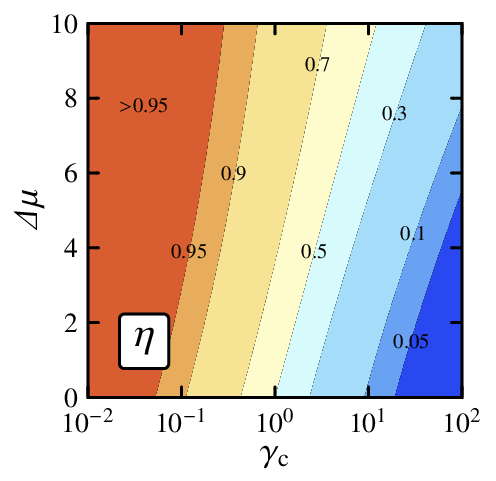}\label{fig1b}}%
\subfigure[]{\includegraphics{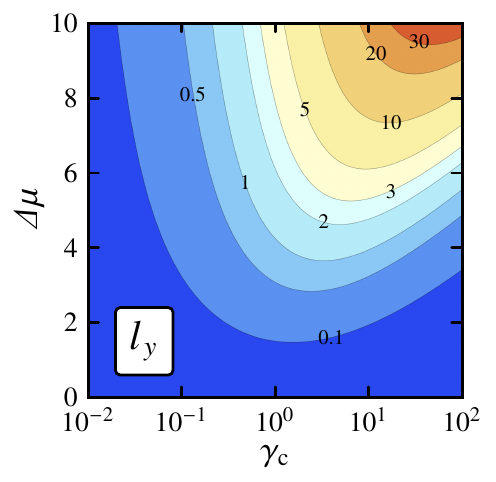}\label{fig1c}}\\
\subfigure[]{\includegraphics{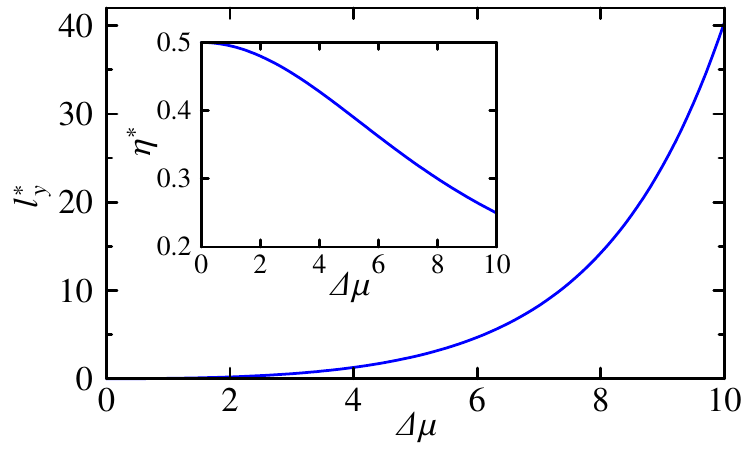}\label{fig1d}}%
\caption{Toy model. (a) Transition rates. (b) Efficiency $\eta=l_y/\sigma$ and (c) learning rate $l_y$ as a function of the external transition rate $\gamma_{\rm c}$ and the chemical potential $\varDelta\mu$.
(d) The maximal learning rate $l_y^*$ and, in the inset, the efficiency at maximum power $\eta^*$. The rates are chosen as $\omega_+=\kappa_+= \exp(\varDelta \mu/2)$ and $\omega_-=\kappa_-= 1$.
The numerical results throughout the paper were obtained by solving the respective master equation numerically.}
\label{fig1}%
\end{figure}

We start with the simplest thermodynamically consistent model for which the inequality (\ref{mainine}) can be studied \cite{bara13b}, see Fig. \ref{fig1a}.
The internal process corresponds to a protein inside the cell Y, which can be activated and deactivated by phosphorylation and dephosphorylation reactions. The chemical potential 
difference driving the reactions is $\varDelta \mu\equiv \mu_{\rm ATP}-\mu_{\rm ADP}-\mu_{\rm P}\ge 0$. These reactions are represented by 
\begin{equation}
 {\rm Y}+{\rm ATP}\xrightleftharpoons[\kappa_-]{\kappa_+} {\rm Y}^*+{\rm ADP} \xrightleftharpoons[\omega_-]{\omega_+} {\rm Y}+{\rm ADP}+{\rm P}_i,
\label{eqreaction} 
\end{equation}    
where $\kappa_+$ is the phosphorylation rate and $\omega_+$ is the dephosphorylation rate, with $\kappa_-$ and $\omega_-$ representing the rates 
of the respective reversed reactions. The reaction rates are related to the chemical potential difference through $\varDelta \mu= \ln[\kappa_+\omega_+/(\kappa_-\omega_-)]$, where we set $k_{\rm B}T\equiv 1$ throughout the paper. 

The external process corresponds to the external ligand concentration which jumps between low and high concentration with rate $\gamma_{\rm c}$.
The internal reaction rates  depend on the external concentration in the following way. It is assumed that if the concentration is high
the receptors in the cell surface are occupied by a ligand and in the inactive state, whereas if the concentration is low they are unbound and in the active state. Moreover,
an active receptor enhances the phosphorylation rate of the internal chemical reaction. More precisely, we consider that if the concentration is high, implying inactive receptors,
only dephosphorylation occurs and if the concentration is low only phosphorylation takes place. This leads to the four state model represented in Fig. \ref{fig1a}. Note that a full model
with both chemical reactions occurring for both low and high concentration would have two links, representing the different chemical reactions, for the
vertical transitions in Fig. \ref{fig1a} \cite{bara13b}. This model is a quite simplified description of the {\mysl E. coli} chemotaxis signaling network \cite{bial12}, with the internal 
protein corresponding to the CheY protein,
which binds to the flagellar motor in its phosphorylated form, inducing tumbling. See \cite{meht12} for a related model where the number of phosphorylated  internal proteins can be large.

The stationary probability current $J= \gamma_{\rm c}(P_1-P_3)$, where $P_1$ and $P_3$ are the stationary probabilities of the states shown in Fig. \ref{fig1a},
can be easily calculated and is given by
\begin{equation}
J= \frac{\gamma_{\rm c}}{2} \frac{\kappa_+\omega_+-\kappa_-\omega_-}{\gamma_{\rm c}[\kappa_++\kappa_-+\omega_++\omega_-]+(\kappa_++\kappa_-)(\omega_++\omega_-)}.
\label{currsimp}
\end{equation}
The thermodynamic entropy production $\sigma$ equals the rate of ATP consumption, i.e.,  
\begin{equation}
\sigma=  J\varDelta \mu.
\end{equation}
For a nonzero learning rate the cell must consume ATP and   
\begin{equation}
l_y= J f,
\end{equation}
where $f=\ln [P_1P_4/(P_3P_2)]$, is bounded by the rate of ATP consumption. This model shows tight coupling since $l_y$ and $\sigma$ are proportional to the same probability current $J$.
It allows for an analogy with molecular motors \cite{seif12}. Whereas molecular motors transform chemical energy obtained from ATP hydrolysis into mechanical work, in our model
ATP is consumed so that the internal protein can learn about the external process at rate $l_y$.  

There are two special limits related to the time-scales of the external process. For $\gamma_{\rm c}\ll \kappa_\pm,\omega_\pm$ the current (\ref{currsimp}) goes to zero and $f\to \varDelta \mu$, implying maximal efficiency. 
Therefore, this limit of zero learning rate and efficiency one corresponds to an adiabatic case.  
Second, for $\gamma_{\rm c}\gg \kappa_\pm,\omega_\pm$ the current tends to its maximal value for fixed $\varDelta\mu$
\begin{equation}
J= \frac{1}{2} \frac{\kappa_+\omega_+-\kappa_-\omega_-}{\kappa_++\kappa_-+\omega_++\omega_-},
\end{equation} 
but $f\to 0$, implying $l_y=0$. This limit is very inefficient, with zero learning rate and maximal ATP consumption for fixed $\varDelta\mu$.
These two limits are represented in the efficiency diagram plotted in Fig. \ref{fig1b}.
Comparing again with a molecular motor, the second limit corresponds to the case where the mechanical force goes to zero, leading to a high velocity of the motor but no mechanical work extraction. The first case, $\gamma_{\rm c}\ll \kappa_\pm,\omega_\pm$,
is related to the stall force case, where the mechanical force equals the chemical potential difference driving the motor and the velocity tends to zero with efficiency approaching one.

From these two limiting cases it is clear that, for fixed $\varDelta \mu$, there is an optimal time-scale $\gamma_{\rm c}^*(\varDelta \mu)$ for which the learning rate $l_y$ should be maximal. This can be seen
in Fig \ref{fig1c}. The maximum learning rate $l_y^*(\varDelta \mu)\equiv l_y(\varDelta \mu,\gamma^*_c)$ increases with $\varDelta\mu$, while the efficiency at maximum power
$\eta^*\equiv l_y^*/\sigma^*$ decreases with $\varDelta\mu$, where $\sigma^*(\varDelta\mu)= \sigma(\varDelta \mu,\gamma^*_c)$, as plotted in Fig. \ref{fig1d}. 
Furthermore, the efficiency at maximum power is maximal near equilibrium where it  tends to $1/2$, a well known result from linear response theory for tightly coupled heat engines \cite{vdb05} 
and molecular motors \cite{seif11a}.  

The entropy production corresponding to the rate of ATP consumed inside the cell is specific to this model. As we show next the entropy production can also arise from work done
by the external process.

\section{Single Receptor Model}
\label{sec4}

{\it E. coli} receptors sit at its membrane and external ligands molecules can bind to them. The kinase CheA is connected to
the receptor through a protein CheW and its activity is influenced by the binding of external ligands. If CheA is in the active form, it acts as an enzyme of
the phosphorylation reaction of the protein CheY \cite{bial12}. Here we consider a MWC model for a single receptor accounting for the indirect regulation of the 
kinase activity by the binding events \cite{skog13,mell07}. Even though this single receptor model contains indirect regulation, it does lack cooperativity, which is the other
key concept in MWC models \cite{marz13}. To study cooperativity more binding sites are needed which corresponds to a straightforward extension of the model. As we do not
investigate cooperativity effects but rather the relation between learning rates and energy consumption, it is more convenient to keep a simpler 
model with a reduced number of states.   

\begin{figure}
\centering%
\subfigure[]{\raisebox{4mm}{\includegraphics{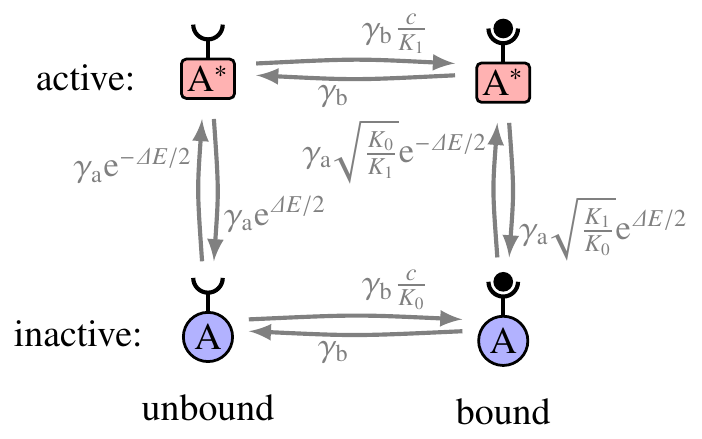}}%
\label{fig2a}}%
\subfigure[]{\includegraphics{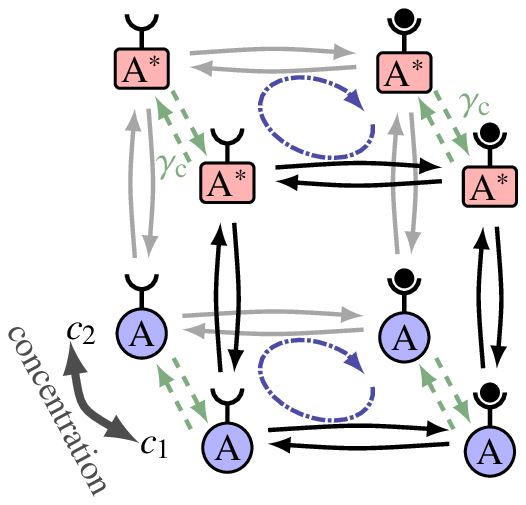}%
\label{fig2b}}%
\caption{ Single receptor model. (a) Transition rates for the four-state single receptor model corresponding to the internal process. 
(b) The full model including the external transitions between concentrations $c_1$ and $c_2$, with the external transitions represented by the green dashed arrows. The blue dash-dotted
cycles contribute to the entropy production in Eq. (\ref{chemicalw}).}
\label{fig2}
\end{figure}%

More precisely, the internal process $y=(a,b)$ corresponds to the following four-state model, see Fig. \ref{fig2a}. 
The receptor can be either bound $b=1$ (occupied by an external ligand) or unbound $b=0$. The free energy difference between the
bound and unbound states is given by $F_{\rm b}-F_{\rm u}= \ln(K/c)$, where $K$ is the dissociation constant and $c$ the external ligand concentration. The quantities $F_{\rm b}$ and $F_{\rm u}$ represent the free energies of
the receptor together with the external solution: in $F_{\rm b}-F_{\rm u}$ the term $\ln K$ is related to a change in the free energy of the receptor and $\ln c$ is the chemical potential of the particle taken 
from the solution in a binding event \cite{seif11}. Moreover, the kinase attached to the receptor can be either inactive $a=0$ or active $a=1$. A conformational change in the receptor is assumed to be an equilibrium 
process with the free energy difference between active and inactive given by $\varDelta E$.

The interaction between the receptor and the enzyme attached to it is reflected in the dissociation constant 
depending on $a$, being $K_0$ for $a=0$ and $K_1$ for $a=1$. It is assumed that $K_1>K_0$, with the activity increasing the dissociation constant. 
Combining these parameters, the free energy of an internal state $(a,b)$ can be written as
\begin{equation}
F(a,b,c)= a \varDelta E
- b\ln(c/K_{a}).
\end{equation}
The transition rates of this four-state model $w_{(ab)(a'b')}$, from state $(a,b)$ to state $(a',b')$, have to respect the detailed balance relation $\ln[w_{(ab)(a'b')}/w_{(a'b')(ab)}]= F(a,b,c)-F(a',b',c)$. The four-state model 
corresponding to the internal process is represented in Fig. \ref{fig2a}, where $\gamma_{\rm a}$ and $\gamma_{\rm b}$ set the time-scale of the conformational changes and binding events, respectively.
The time-scale of the latter is assumed to be much smaller than the time-scale of the conformational changes, i.e., $\gamma_{\rm b}\gg \gamma_{\rm a}$.  
As this is an equilibrium model, the cycle affinity of the four-state cycle is zero.

The external process corresponds to the external ligand concentration jumping between the values $c_1$ and $c_2$ at rate $\gamma_{\rm c}$. The resulting full eight-state model is shown in Fig. \ref{fig2b}. Denoting the stationary probability of
the state $(x;y)= (c;a,b)$ by $P_{a,b}^c$, and using Schnakenberg's formula \cite{schn76}, where the terms in the entropy production are cycle affinities multiplying probability currents,  
it is possible to write the entropy production (\ref{thermoent}) in the form
\begin{equation}
\sigma= \gamma_{\rm c}\sum_a(P_{a,1}^{c_2}- P_{a,1}^{c_1})\ln \frac{c_2}{c_1}.
\label{chemicalw}
\end{equation}    
This entropy production arises from cycles, as indicated  in Fig. \ref{fig2b}, where a ligand molecule is taken from the solution at concentration $c_2$ and released in the solution at concentration $c_1$,
corresponding to a chemical potential difference $\ln(c_2/c_1)$. Therefore, $\sigma$ is the chemical work that is done by the external process. In contrast to the first toy model, where the cell consumes ATP, 
the work done by the external process compensates for a nonzero learning rate $l_{\rm ab}$, which follows from (\ref{learnrate}) as
\begin{equation}
l_{\rm ab}=\gamma_{\rm c}\sum_{a,b}(P^{c_2}_{a,b}-P^{c_1}_{a,b})\ln\left(\frac{P^{c_2}_{a,b}}{P^{c_1}_{a,b}}\right).
\label{eq:hx_8state}
\end{equation} 
We also consider the coarse-grained rate at which $a$ learns about $c$, which is obtained from (\ref{learnratecg}) as
\begin{equation}
l_{\rm a}=\gamma_{\rm c}\sum_{a}(P_a^{c_2}-P_a^{c_1})\ln \frac{P_a^{c_2}}{P_{a}^{c_1}},
\end{equation}
where $P_a^c=\sum_b P_{a,b}^c$. The rate $l_{\rm a}$ characterizes how much the internal kinase CheA learns about the external process. This is the
central quantity assuming that other chemical reactions inside the cell that are influenced by the activity can only learn as much as $l_{\rm a}$.

\begin{figure}
 \centering
 \subfigure[]{\includegraphics[width=0.333\textwidth]{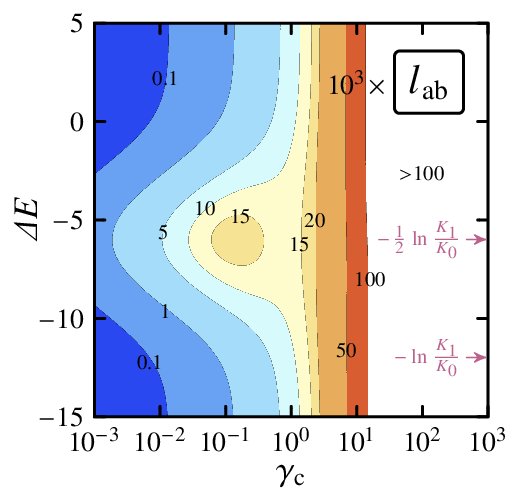}\label{fig4a}}%
 \subfigure[]{\includegraphics[width=0.333\textwidth]{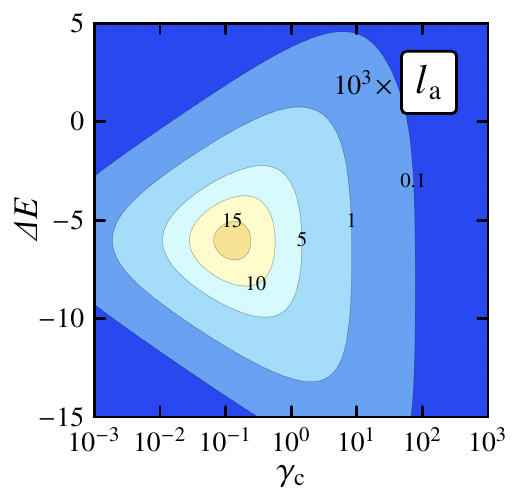}\label{fig4b}}%
 \subfigure[]{\includegraphics[width=0.333\textwidth]{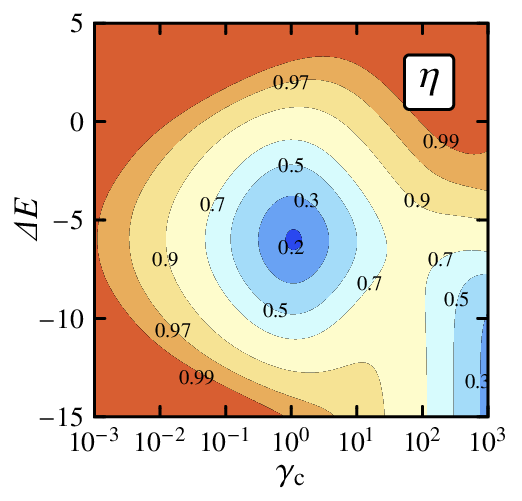}\label{fig4c}}
 \caption{Numerical results for the single receptor model. (a) Learning rate $l_{\rm ab}\times 10^3$, (b) coarse-grained leaning rate $l_{\rm a}\times 10^3$ and (c) efficiency $\eta=l_{\rm ab}/\sigma$ as functions of
the external transition rate $\gamma_{\rm c}$ and  the conformational free energy difference $\varDelta E$. The other parameters are set to $K_1=(K_0)^{-1}= 400$, $\gamma_{\rm a}=1$, $\gamma_{\rm b}= 10^3$, $c_2=(c_1)^{-1}=3$.}
\label{fig4}
\end{figure}
\begin{figure}
\centering
 \subfigure[]{\includegraphics{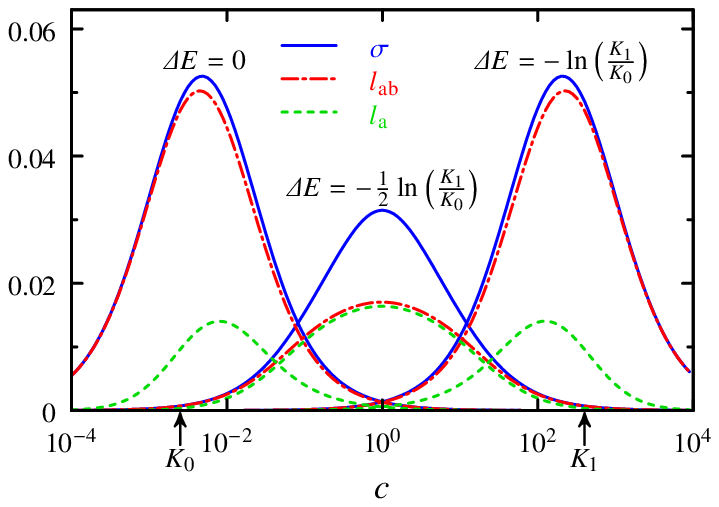}\label{fig5a}}
 \subfigure[]{\includegraphics{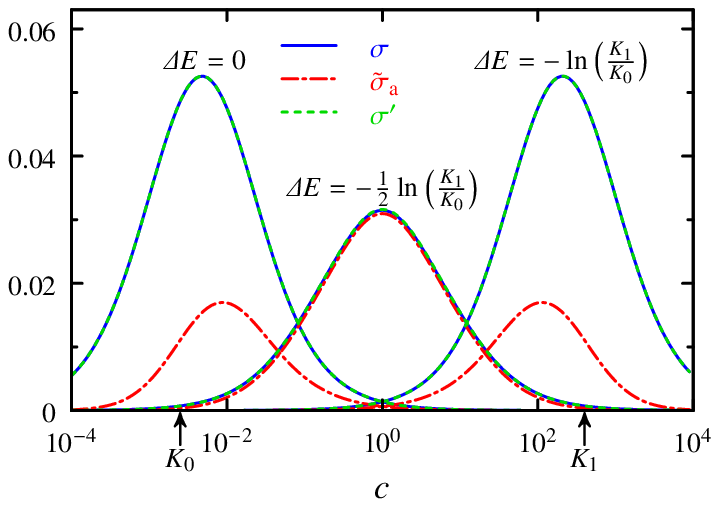}\label{fig5b}}
\vspace{-2mm}
\caption{Numerical results for the single receptor model. 
(a) The learning rate $l_{\rm ab}$, the chemical work $\sigma$ and the coarse-grained learning rate $l_{\rm a}$. (b) The chemical work $\sigma$, the lower bound on the chemical work $\tilde{\sigma}_{\rm a}$, 
and the chemical work with time-scale separation $\sigma'$. The variable $c$ parametrizes the concentration as $c_2=3c$ and $c_1=c/3$. The other parameters are set to $K_1=(K_0)^{-1}= 400$, $\gamma_{\rm b}=10^3$, 
$\gamma_{\rm a}=1$, and $\gamma_{\rm c}=10^{-1}$.}   
\label{fig5}
\end{figure}

Numerical results for this model with the choice of transition rates shown in Fig. \ref{fig2}  are shown in Fig. \ref{fig4}. Comparing Figs. \ref{fig4a} and \ref{fig4b},
we see that the coarse-grained learning rate $l_{\rm a}$ is close to $l_{\rm ab}$ if $\gamma_{\rm c}$ is small. As $\gamma_{\rm c}$ increases, the difference between both 
quantities increases. More precisely, if the external environment is considerably faster then the internal variable $a$ ($\gamma_{\rm c}\gg\gamma_{\rm a}$), then $a$ cannot track concentration changes ($l_{\rm a}\to0$). However, 
the learning rate $l_{\rm ab}$ is still nonzero, with its main contribution coming from the variable $b$. 

The efficiency plot in Fig. \ref{fig4c} demonstrates that if the external environment is very slow $\gamma_{\rm c}\ll \gamma_{\rm a}$ the model reaches an adiabatic limit with $\eta\to 1$ and $l_{\rm ab}\to 0$. 
For $\gamma_{\rm c}\simeq 0.1$ there is a region of lower efficiency $\eta\simeq 0.5$ and high learning rates $l_{\rm ab}\simeq l_{\rm a}> 0.015$. 
If $\gamma_{\rm c}$ is further increased beyond the values displayed in Fig. \ref{fig4c}, the efficiency decays, going to zero for $\gamma_{\rm c}\gg \gamma_{\rm b}$. In this case, not 
even $b$ can track the fast external changes.

Choosing the external concentrations as $c_1= c/3$ and $c_2=3c$, in Fig. \ref{fig5a} we plot $l_{\rm ab}$ and $\sigma$ as a function of $c$ for different values of the conformational free energy difference $\varDelta E$. Besides 
the simple illustration that the learning rate is bounded by the chemical work, these graphs demonstrate that for every $c$ there is an optimal value of $\varDelta E$ that maximizes the learning rate $l_{\rm ab}$.    
This is also the case for the coarse-grained learning rate $l_{\rm a}$. Comparing $l_{\rm ab}$ with $l_{\rm a}$, we see that the contribution of the variable $b$ to $l_{\rm ab}$ is more substantial for
very low and very high $c$. However, for $K_0\ll c\ll K_1$  the correlation between bound (unbound) and inactive (active) is high leading to $l_{\rm ab}\simeq l_{\rm a}$.

\begin{figure}%
 \centering%
 \subfigure[]{\raisebox{8.55mm}{\includegraphics{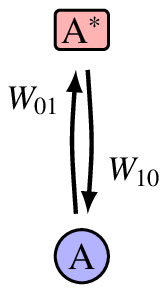}}\label{fig3a}}\quad
  \subfigure[]{\includegraphics{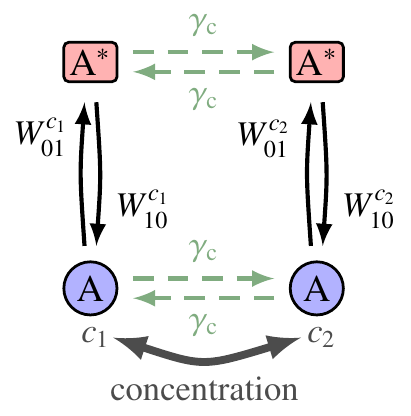}\label{fig3b}}%
\caption{Coarse-grained single receptor model. (a) Coarse-grained internal process with the transition rates defined in (\ref{cgrates}). (b) Full coarse-grained model with the external jumps. Note that the vertical transition rates, which are also obtained from (\ref{cgrates}),
contain an explicit dependence on the concentration for the full model.}
\label{fig3}%
\end{figure}%
Since binding events are faster than the conformational changes, it is possible to integrate out the variable $b$ \cite{skog13,tu08b}. 
In this case, the internal process reduces to the two-state model of Fig \ref{fig3a}.
In the limit $\gamma_{\rm a}/\gamma_{\rm b}\to 0$ the coarse-grained transition rates become \cite{espo12}
\begin{equation}
W_{aa'}=\sum_b w_{(ab)(a'b)}\frac{(c/K_a)^b}{1+c/K_a},
\label{cgrates}
\end{equation}   
where $(c/K_a)^b/(1+c/K_a)$ corresponds to the stationary conditional probability $P(b|a)$ of the full internal process. The effective free energy difference resulting from a conformational change from $a=0$ to $a=1$ then reads
\begin{equation}
\varDelta F(c)= \ln\frac{W_{10}}{W_{01}}= \varDelta E+ \ln\left(\frac{1+\frac{c}{K_0}}{1+\frac{c}{K_1}}\right).    
\label{coarsefree}
\end{equation}     
The full system, including the concentration jumps, is the four-state model shown in Fig. \ref{fig3b}.
The entropy production for this model is the coarse-grained entropy production (\ref{thermoentcg}), which reads
\begin{equation}
\tilde{\sigma}_{\rm a}= \gamma_{\rm c}(P_0^{c_2}-P_0^{c_1})\ln\left[\left(\frac{1+\frac{c_2}{K_0}}{1+\frac{c_2}{K_1}}\right)\left(\frac{1+\frac{c_1}{K_1}}{1+\frac{c_1}{K_0}}\right)\right],
\label{coarw}
\end{equation}
where $P_a^c$ denotes the stationary probability of the coarse-grained model. This quantity provides a lower bound on the full chemical work (\ref{chemicalw}), which in the limit  $\gamma_{\rm a}/\gamma_{\rm b}\to 0$ becomes
\begin{equation}
\sigma'= \gamma_{\rm c} \sum_a\left(P_a^{c_2}\frac{c_2}{K_a+c_2}-P_a^{c_1}\frac{c_1}{K_a+c_1}\right)\ln \frac{c_2}{c_1}.
\label{fullw}
\end{equation}
Therefore, with this time-scale separation the full chemical work can be calculated with the coarse-grained model by using this formula. In Fig. \ref{fig5b} we compare $\sigma$ for $\gamma_{\rm a}/\gamma_{\rm b}= 10^{-3}$, $\sigma'$, and
$\tilde{\sigma}_{\rm a}$. The important result shown in this figure is that for $K_0\ll c\ll K_1$ the lower bound $\tilde{\sigma}_{\rm a}$ is close to the full chemical work $\sigma'$. In this case, $\tilde{\sigma}_{\rm a}\simeq \sigma'$ since
$c/(K_0+c)\simeq 1$, $c/(K_1+c)\simeq 0$, and the affinity in (\ref{coarw}) becomes $\ln(c_2/c_1)$.

In principle, the conformational change of the receptor could also involve ATP consumption, which would lead to an internal process breaking detailed balance even without the external jumps \cite{skog13}. 
In this case, the entropy production of the full model would involve both energy consumed inside the cell and work done by the external process.

\section{Model with adaptation}
\label{sec5}
In this section, we study a more complete model 
also including adaptation, where the free energy difference arising from  conformational changes for the activity is taken from the coarse-grained model discussed in the previous section.    
\subsection{Model definition}

For adaptation, besides the kinase activity $a$, the internal process must include the methylation level $m$ \cite{lan12}. 
As in the previous model, the activity $a$ takes the values $a=0$ if CheA is  inactive and $a=1$ if it is active. 
The average value of $a$ is assumed to be  around $1/2$, independent of $c$. Whereas $a$ quickly responds to a change in $c$ at a time-scale $\gamma_{\rm a}^{-1}$, the methylation level  $m=0,1,2,3,4$ guarantees 
adaption by returning the average activity to $1/2$ at a time-scale $\gamma_{\rm m}^{-1}\gg \gamma_{\rm a}^{-1}$. More specifically, 
a step decrease in $c$ leads to a fast increase in $a$. This change in $a$ generates a slow decrease in the methylation level $m$ which acts back on the activity slowly decreasing it to $1/2$.

\begin{figure}
\centering
\subfigure[]{\includegraphics{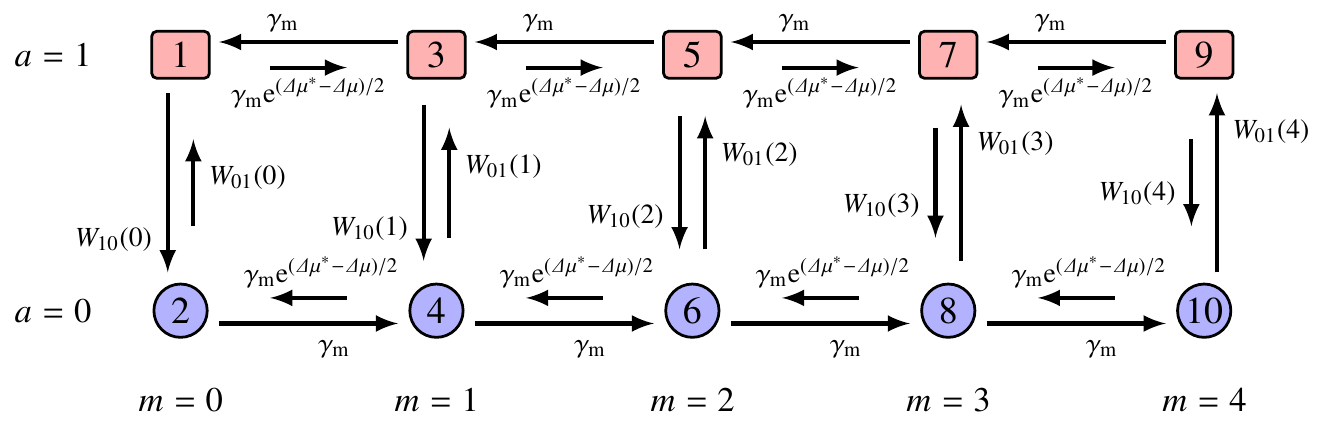}\label{fig6a}}
\subfigure[]{\includegraphics{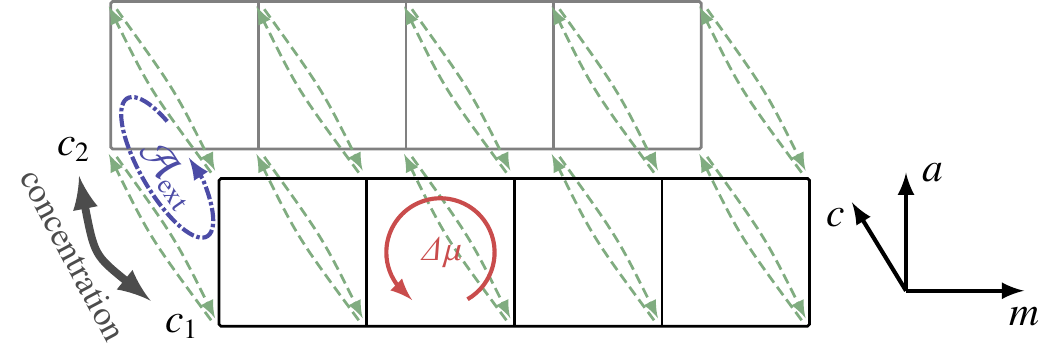}\label{fig6b}}
\caption{Model with adaptation. (a) Internal process for the model with adaptation. The transition rates for the conformational changes are obtained from equation (\ref{cgrates}) with the choice represented in Fig. \ref{fig2a}. Note that in this case
$\Delta E$ depends on $m$ as given by (\ref{dEdef}). (b) The full model with the external jumps. The internal cycles are related to an affinity $\varDelta \mu$ and the external cycles to an affinity $\mathcal{A}_{\rm ext}$, which is the logarithm in (\ref{sigmac}).}
\label{fig6}
\end{figure}

The free energy difference for the conformational change of the activity $a$ is obtained from the coarse-grained  MWC model expression (\ref{coarsefree}). Explicitly, 
\begin{equation}
\varDelta F(m,c)=\varDelta E(m)+\ln\left(\frac{1+\frac{c}{K_0}}{1+\frac{c}{K_1}}\right),
\end{equation}
where the free energy dependence on the methylation level is
\begin{equation}
\varDelta E(m)\equiv -\frac{m}{4}\ln\frac{K_1}{K_0}.
\label{dEdef}
\end{equation}
This choice for the free energy difference leads to $\varDelta F(4,\infty)=\varDelta F(0,0)=0$.
For simplicity, we assume 
\begin{equation}
F(a,m,c)= a\varDelta F(m,c).
\end{equation}

Two different chemical reactions control the methylation level. If $a=0$, the receptor can be methylated with the reaction
\begin{equation}
[m]_0+{\rm SAM} \rightleftharpoons [m+1]_0+{\rm SAH},
 \label{eq:demethylation}
\end{equation}
where the subscript indicate the value of $a$. Here SAM represents a S-Adenosyl methionine molecule and SAH represents a S-Adenosyl-L-homocysteine molecule.
The free energy difference of the reaction is $\mu_{{\rm SAM}}-\mu_{{\rm SAH}}$.
Likewise, if $a=1$,  the receptor can be demethylated with the reaction
\begin{equation}
[m+1]_1+{\rm H_2O} \rightleftharpoons  [m]_1+{\rm CH_3OH},
 \label{eq:methylation}
\end{equation}
for which the free energy difference is 
\begin{equation}
\mu_{{\rm \rm H_2O}}-\mu_{{\rm CH_3OH}}+\varDelta F(m+1,c)-\varDelta F(m,c)=\mu_{{\rm H_2O}}-\mu_{{\rm CH_3OH}}- \frac{1}{4}\ln\frac{K_1}{K_0}.
\end{equation}
The chemical potential difference 
\begin{equation}
\varDelta\mu=\mu_{{\rm SAM}}+\mu_{{\rm H_2O}}-\mu_{{\rm SAH}}-\mu_{{\rm CH_3OH}}
\end{equation} 
then drives the internal process out of equilibrium, corresponding to the affinity of the internal cycles shown in Fig. \ref{fig6}, where our choice for the transition rates for the 
internal process is displayed. We note that if $\varDelta\mu=0$ the internal process is in equilibrium. 
For $0<\varDelta\mu< \varDelta \mu^*\equiv(1/4)\ln(K_1/K_0)$ the system dissipates but there is no adaptation: $m$ increases $a$ instead of repressing it. Adaptation happens
only if $\varDelta \mu$ overcomes the conformational free energy difference in a internal cycle $\varDelta \mu^*$.

The variable $y=(a,m)$ defines the internal process, the external process $x$ is again the fluctuating concentration of external ligand, which jumps with rate $\gamma_{\rm c}$ between  $c_1$ and $c_2$. 
The full model thus has $20$ states, as shown in Fig. \ref{fig6b}. In general, its transition rates are denoted by  $w_{(a_1,m_1)(a_2,m_2)}^c$ and the stationary probability by $P_{a,m}^{c}$. 
Moreover, the internal probability current is 
\begin{equation}
J^{c}_{(0,m)(0,m+1)}\equiv P_{0,m}^c w_{(0,m)(0,m+1)}^c-P_{0,m+1}^c w_{(0,m+1)(0,m)}^c,
\end{equation}
 and the marginal probability is 
\begin{equation}
P_a^c\equiv\sum_m P_{a,m}^{c}.
\label{marg}
\end{equation}

\subsection{Chemical work, SAM consumption and learning rate}

Using Schnakenberg's network theory \cite{schn76}, the entropy production (\ref{thermoent}) can be conveniently written as
\begin{equation}
\sigma= \sigma_{\textrm{int}}+\sigma_{\textrm{ext}},
\end{equation}
where  
\begin{equation}
\sigma_{\rm int}\equiv \sum_{m=0}^3 \left[J^{c_1}_{(0,m)(0,m+1)}+J^{c_2}_{(0,m)(0,m+1)}\right]\varDelta\mu
\label{sigmamu}
\end{equation}
corresponds to the consumption of SAM inside the cell and
\begin{equation}
\sigma_{\rm ext}\equiv \gamma_{\rm c}(P_{0}^{c_2}-P_{0}^{c_1})\ln\left[\left(\frac{1+\frac{c_2}{K_0}}{1+\frac{c_2}{K_1}}\right)\left(\frac{1+\frac{c_1}{K_1}}{1+\frac{c_1}{K_0}}\right)\right]
\label{sigmac}
\end{equation}
is a lower bound on the chemical work done by the external process, as explained in the previous section. 
The term $\sigma_{\rm int}$ is related to the internal cycles in Fig. \ref{fig6b}. 
In principle, the full chemical work can only be obtained if we consider a process with $40$ states, for which
the variable $b$ is not integrated out. However, if there is time-scale separation ($b$ faster than $a$),
we can adapt equation (\ref{fullw}) to include methylation levels.  For the present model, the full work done by the external process then reads  
\begin{equation}
\sigma_{\rm ext}'= \gamma_{\rm c}\sum_a\left(P_a^{c_2}\frac{c_2}{K_a+c_2}-P_a^{c_1}\frac{c_1}{K_a+c_1}\right)\ln\frac{c_2}{c_1},
\label{sigmactot}
\end{equation}
where, contrary to (\ref{fullw}), $P_a^{c_2}$ is now the marginal probability (\ref{marg}). Furthermore,
the learning rate (\ref{learnrate}), becomes 
\begin{equation}
l_{\rm am}= \gamma_{\rm c}\sum_{a,m} (P_{a,m}^{c_2}-P_{a,m}^{c_1})\ln\frac{P_{a,m}^{c_2}}{P_{a,m}^{c_1}}.
\label{lamnat}
\end{equation}

\begin{figure}
\centering
\subfigure[]{\includegraphics{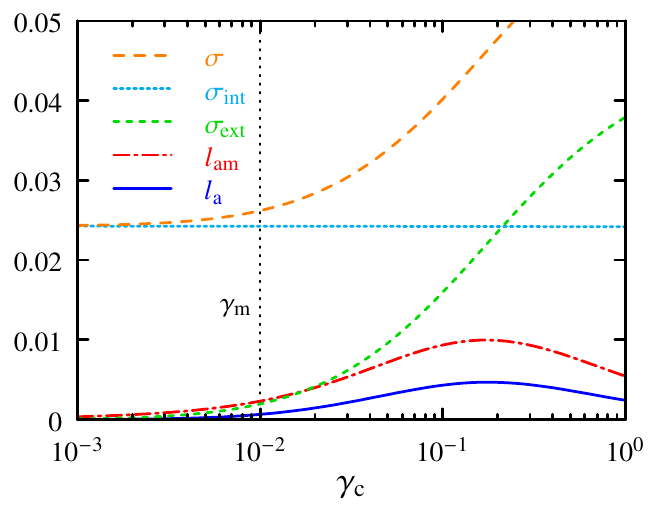}\label{fig7a}}%
\subfigure[]{\includegraphics{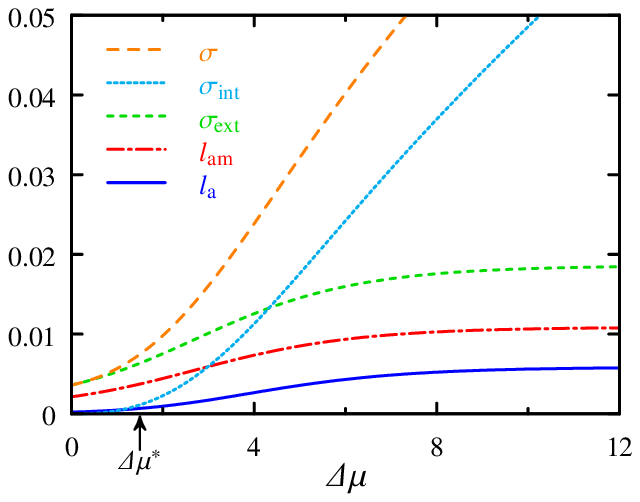}\label{fig7b}}\\
\subfigure[]{\includegraphics{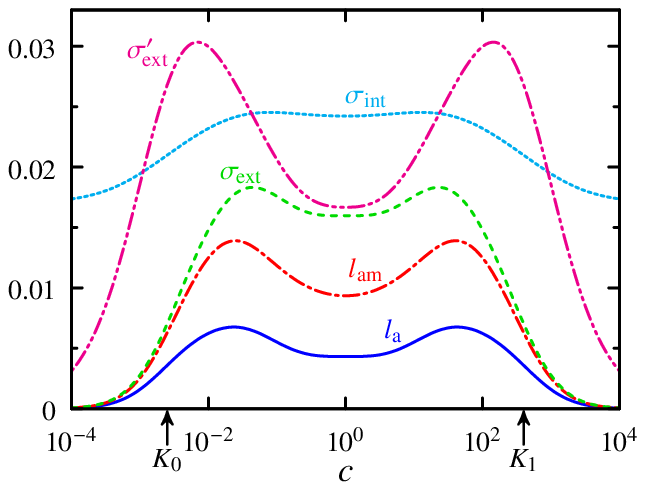}\label{fig7c}}%
\subfigure[]{\includegraphics{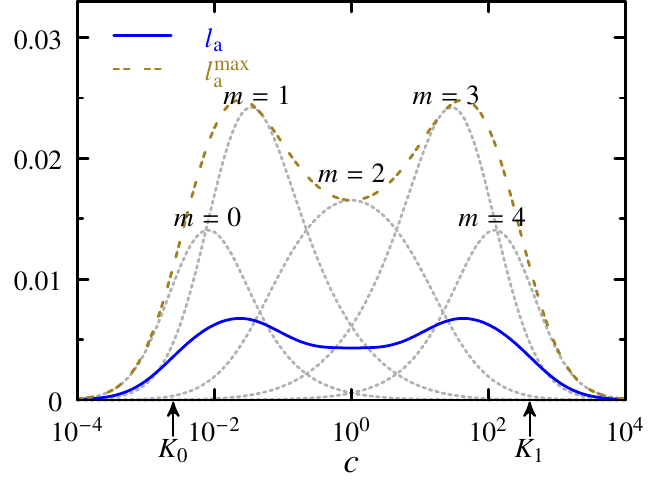}\label{fig7d}}
\caption{Numerical results for the model with adaptation. 
The contributions to the entropy production and the learning rates as a function of (a) the jumping rate of the external concentration $\gamma_{\rm c}$, (b) the chemical potential difference $\varDelta \mu$, and (c) the parameter 
$c$, which define the concentrations through $c_1=c/3$ and $c_2=3c$. (d) The coarse-grained learning rate $l_{\rm a}$ for the model with adaptation compared to learning rates for the model without adaptation plotted in Fig. \ref{fig5a} for
different values of the free energy difference $\varDelta E$, represented by the gray dotted lines. The values of  $\varDelta E$ correspond to $\varDelta E(m)$ in (\ref{dEdef}) for methylation level $m=0,1,2,3,4$.
The quantity $l_{\rm a}^{\rm max}$ is the maximal achievable  learning rate by fine-tuning $\varDelta E$ with fixed $c$ in the model without adaptation. The parameters are set to 
$K_1=(K_0)^{-1}= 400$; $\gamma_{\rm a}=1$; $\gamma_{\rm m}=10^{-2}$; $c=1$ in (a) and (b); $\gamma_{\rm c}=10^{-1}$ in (b), (c) and (d); $\varDelta\mu=6$ in (a), (c) and (d).}
\label{fig7}
\end{figure}

Numerical results for this model are shown in Fig. \ref{fig7}. From the plot of the thermodynamic entropy production and the learning rate as a function of
$\gamma_{\rm c}$ in Fig. \ref{fig7a}, we see that if the external process is very slow the dissipation inside the cell due to SAM consumption $\sigma_{\rm int}$ is much larger than the learning rate.
Therefore, if the bacterium swims in an environment with $\gamma_{\rm c}\ll\gamma_{\rm a}$, it will dissipate much more than it learns about the environment. The SAM consumption rate is nearly independent
of $\gamma_{\rm c}$, being related to probability currents in the internal cycles as expressed in Eq. (\ref{sigmamu}), while $\sigma_{\rm ext}$ grows with $\gamma_{\rm c}$. The learning rate $l_{\rm am}$ reaches
a maximum, similar to the behavior observed with the toy model from Sec. \ref{sec3} and the MWC model from Sec. \ref{sec4}. 

Fig. \ref{fig7b} demonstrates that  the learning rate $l_{\rm am}$ can be larger than the rate of SAM consumption $\sigma_{\rm int}$ inside the cell since the work done
by the external process also contributes to the cost of the learning rate. The internal dissipation $\sigma_{\rm int}$ grows with $\varDelta \mu$, while the learning rate $l_{\rm am}$ saturates for high $\varDelta \mu$.

In adaptation, the fast variable $a$ learns about the changes in external concentration whereas the function of the slow variable $m$ is to appropriately regulate the conformational 
free energy difference $\varDelta E(m)$. With such a perspective, it is more meaningful to look at the coarse-grained learning rate $l_{\rm a}$, which from (\ref{learnratecg}) follows as
\begin{equation}
l_{\rm a}= \gamma_{\rm c}\sum_a (P_a^{c_2}-P_a^{c_1})\ln\frac{P_a^{c_2}}{P_a^{c_1}}.
\label{lanat}
\end{equation}
The relation between the learning rates and the external concentration is plotted in  Fig. \ref{fig7c}. The external concentration are fixed  to $c_1= c/3$ and $c_2=3c$, with $c$ as  free parameter.
The learning rates $l_{\rm am}$ and $l_{\rm a}$ approach zero if $c$ increases (decreases) beyond $K_1$ ($K_0$). Moreover, in Fig.  \ref{fig7c}
we see that the difference between $\sigma_{\rm ext}$ and $\sigma_{\rm ext}'$ is small in the region $K_0\ll c\ll K_1$.  

An interesting result is obtained in Fig. \ref{fig7d} where we compare $l_{\rm a}$ with the coarse-grained learning rate obtained for the MWC model without adaptation plotted in Fig. \ref{fig5a}. For fixed $\varDelta E$, i.e., without adaptation,
the maximal learning rate $l_{\rm a}^{\rm max}$
achieved at any given value of $c$ by fine-tuning $\varDelta E$ represents the maximum the kinase dynamics can learn about the environment for given $c$. 
As demonstrated in Fig. \ref{fig7d}, the learning rate $l_{\rm a}$ for the model with adaptation is below $l_{\rm a}^{\rm max}$. The advantage of adaptation, however, is that it increases the 
region at which $l_{\rm a}$ is non-negligible (see comparison between the blue solid and gray dotted curves in Fig. \ref{fig7d}), 
which is roughly $K_0<c<K_1$. Similarly, it has been found that adaptation maintains high sensitivity over a wide concentration range \cite{mell07} (see \cite{cela10} for a different perspective).

A closely related study of the present model for {\mysl E. coli}  adaptation has been recently performed by Sartori et al. \cite{sart14}. In contrast to our model, in their work the external concentration is suddenly
changed from an initial to a final value with some preassigned probability. 
The mutual information between the internal system $(a,m)$ and the final concentration, which is related to measuring the final concentration, is analogous to our learning rate $l_{\rm am}$. In \cite{sart14} this mutual 
information is shown to increase with time, reaching its maximal value at a time of the order of the methylation time-scale. This is in agreement with our result that the learning rate is much smaller than the
internal dissipation for $\gamma_{\rm c}<\gamma_{\rm m}$: after a time of order $(\gamma_{\rm m})^{-1}$ the correlation with the external concentration is saturated but the cell keeps dissipating.


\section{Conclusion}
\label{sec6}

For Markov processes that can be decomposed into an external process unaffected by an internal process, as
represented by the transition rates in Eq. (\ref{defrates2}), we have defined an entropic rate that is bounded by the thermodynamic 
entropy production and that characterizes how much the internal process learns about the external process. 
This learning rate allows for the definition of an informational efficiency that, as demonstrated with three 
different models related to {\mysl E. coli} sensory network, can be used to study the thermodynamics of cellular information processing.

We have analyzed a simple toy model  of an internal protein tracking an external concentration, for which
the learning rate is bounded by the rate of ATP consumption inside the cell. This thermodynamically consistent model allows for a comparison with molecular motors,
where the learning rate plays the role of extracted mechanical work.

For an internal process corresponding to an equilibrium MWC model for a single {\mysl E. coli} receptor with no internal dissipation, we have shown that  a nonzero 
learning rate is possible if work done by the external process accounts for it. In this model, the work comes from the chemical potential difference of binding a ligand at concentration $c_1$ and
unbinding it at another concentration $c_2$. If the external medium changes much faster than the activity, the full learning rate is nonzero because the binding and unbinding process, which is assumed to be faster than activity changes, 
can also learn about the external environment. However, the learning rate of the activity alone goes to zero, as the activity cannot track an environment much faster than its own time-scale.

Our framework has also been applied to a model including adaptation. We have shown that a bacterium in an external environment that changes at a time-scale much larger than the time-scale of the activity, 
dissipates through SAM consumption at a rate much higher than the one of learning about the external environment, corresponding to a quite inefficient situation. 
In general, the rate of SAM consumption does not bound the learning rate, which can increase due to work done by the external process. Using a coarse-grained learning rate, we have analyzed
how much the activity alone learns about the external process. We have shown that adaptation  increases the concentration range for which this learning rate is non-negligible.

We have assumed the external process to be an external ligand concentration jumping between two values. 
However, the framework from Sec. \ref{sec2} is also valid for more elaborate external processes. Likewise, more complete internal processes
that include further elements of the {\mysl E. coli} sensory network could be studied in the future. We expect features like
the thermodynamic entropy production having one contribution due to energy dissipation inside the cell and another one due work done by the external process, 
the learning rate being much smaller than the internal dissipation if changes in the external medium are slow, and the learning rate not necessarily being bounded
by the dissipation inside the cell to be relevant also for more complex models. Finally, it would be worthwhile to explore the relation between 
the learning rate studied here and quantities characterizing further aspects of a sensory system, like adaptation error and sensitivity.


\section*{References}

\input{thermocellpaper2.bbl}


\end{document}

%% file: thermocellpaper2.bbl
\providecommand{\newblock}{}